\pdfoutput=1
\documentclass[11pt,twoside,a4paper,cmspaper,final,collab]{cms-tdr}
\def\svnVersion{0e56f79}\def\svnDate{2023/11/20}\def\cmsCernNoTag{CERN-EP-2023-093}\def\cmsCernDate{\today}\def\cmsMessage{Published in the Journal of Instrumentation as \href{http://dx.doi.org/10.1088/1748-0221/18/11/P11017}{\doi{10.1088/1748-0221/18/11/P11017}.}}
\begin{document}\cmsNoteHeader{PRF-22-001}

\newcommand{\abseta}{\ensuremath{\abs{\eta}}\xspace}
\newcommand{\absetatrack}{\ensuremath{\abs{\eta_\text{track}}}\xspace}
\newcommand{\NTS}{\ensuremath{N_\text{TS}}\xspace}
\newcommand{\Npe}{\ensuremath{N_\text{pe}}\xspace}
\newcommand{\ieta}{\ensuremath{i_\eta}\xspace}
\newcommand{\iphi}{\ensuremath{i_\phi}\xspace}

\newcommand{\SOI}{\ensuremath{\text{SOI}}\xspace}
\newcommand{\SOIm}{\ensuremath{\text{SOI}{-1}}\xspace}
\newcommand{\SOIp}{\ensuremath{\text{SOI}{+1}}\xspace}

\newcommand{\ASOI}{\ensuremath{A_{\SOI}\xspace}}
\newcommand{\ASOIm}{\ensuremath{A_{\SOIm}\xspace}}
\newcommand{\ASOIp}{\ensuremath{A_{\SOIp}\xspace}}

\newcommand{\Deltaslew}{\ensuremath{\Delta_\text{slew}}\xspace}

\newcommand{\fC}{\ensuremath{\unit{fC}}\xspace}
\newcommand{\ns}{\ensuremath{\unit{ns}}\xspace}
\newcommand{\m}{\ensuremath{\unit{m}}\xspace}

\newcommand{\tmean}{\ensuremath{\langle t\rangle}\xspace}
\newcommand{\sigmat}{\ensuremath{\sigma_t}\xspace}
\newcommand{\pedmean}{\ensuremath{\langle \text{ped}\rangle}\xspace}
\newcommand{\sigmaped}{\ensuremath{\sigma_\text{ped}}\xspace}
\newcommand{\mimutped}{\ensuremath{m_i(\vec{\mu}, \vec{t}, \text{ped})}\xspace}
\newcommand{\sigmapi}{\ensuremath{\sigma_{\mathrm{p},i}}\xspace}

\newcommand{\Dnoise}{\ensuremath{\mathbf{D}^\text{noise}}\xspace}
\newcommand{\Dpulse}{\ensuremath{\mathbf{D}^\text{pulse}}\xspace}
\newcommand{\Dpulsej}{\ensuremath{\Dpulse_j}\xspace}

\cmsNoteHeader{PRF-22-001}
\title{Performance of the local reconstruction algorithms for the CMS hadron calorimeter with Run 2 data}
\author{The CMS Collaboration}
\date{\today}

\abstract{
A description is presented of the algorithms used to reconstruct energy deposited in the CMS hadron calorimeter during Run 2 (2015--2018) of the LHC. During Run 2, the characteristic bunch-crossing spacing for proton-proton collisions was 25\ns, which resulted in overlapping signals from adjacent crossings. The energy corresponding to a particular bunch crossing of interest is estimated using the known pulse shapes of energy depositions in the calorimeter, which are measured as functions of both energy and time. A variety of algorithms were developed to mitigate the effects of adjacent bunch crossings on local energy reconstruction in the hadron calorimeter in Run 2, and their performance is compared.
}

\hypersetup{
pdfauthor={CMS Collaboration},
pdftitle={Performance of the local reconstruction algorithms for the CMS hadron calorimeter with Run 2 data},
pdfsubject={CMS},
pdfkeywords={CMS, HCAL, Local reconstruction}} 

\maketitle 

\section{Introduction}
The hadron calorimeter (HCAL) plays a central role in the reconstruction of events recorded by the CMS detector~\cite{Chatrchyan:2008aa}. 
Its main purpose is to identify both charged and neutral hadrons and measure their energies, and it serves an important role in identifying leptons and photons. 
Its hermetic design, with geometric coverage up to pseudorapidities of $\abseta=5.2$, and fine lateral segmentation aid in the estimation of missing transverse momentum, \ptmiss.

In August 2015, the CERN LHC began delivering proton-proton ($\Pp\Pp$) collisions with a bunch-crossing spacing as short as 25\ns~\cite{Sirunyan2021}; however, the recorded pulse shapes from energy depositions in the HCAL barrel (HB) and endcap (HE) detectors typically last longer than 25\ns.
Approximately 85--90\% of the integrated energy occurs within a 50\ns window.
The energy deposited in the HB and HE from collisions from nearby bunch crossings, referred to as out-of-time pileup (OOTPU), can spoil the energy estimation from the collision of interest. 
We describe the performance of four different algorithms, referred to as local reconstruction algorithms, used to estimate the energy deposited in the detector elements of the HB and HE during Run 2 of the LHC (2015--2018). 
These four algorithms exhibit different levels of sophistication in their ability to mitigate the effects of OOTPU, as well as varying degrees of computational complexity. 
The tradeoff between the ability of the algorithms to suppress OOTPU and their computational performance is important when deploying them in the high-level trigger (HLT)~\cite{Khachatryan:2212926}, the second level of a two-tier trigger system consisting of a farm of processors running an online version of the full event reconstruction. 
The constraints from the HLT thus guide the design of the algorithms and play a key role in their evolution in both online and offline reconstructions.

The paper is organized as follows:
Section~\ref{sec:detector} describes the CMS detector with a particular emphasis on the HCAL.
Section~\ref{sec:methods} describes the four different local reconstruction algorithms used in the HB and HE, and Section~\ref{sec:performance} evaluates their performance.
This paper concludes with a summary in Section~\ref{sec:summary}.

\section{The CMS detector}\label{sec:detector}
The central feature of the CMS apparatus is a superconducting solenoid of 6\m internal diameter, providing a magnetic field of 3.8\unit{T}.  
Within the solenoid volume are a silicon pixel and strip tracker, a lead tungstate crystal electromagnetic calorimeter (ECAL), and the HB and HE.  
The HB and HE are both brass and scintillator sampling calorimeters with pseudorapidity coverages of approximately $\abseta < 1.3$ and $1.3 < \abseta < 3.0$, respectively.
Muons are detected in gas-ionization chambers embedded in the steel flux-return yoke outside the solenoid.
In addition to the HB and HE, the HCAL subsystem also includes the HCAL outer (HO) detector and HCAL forward (HF) detector. 
The HO is composed of plastic scintillators located between the solenoid and the barrel muon system, and measures the energy from very energetic hadronic showers that are not fully contained within the HB and punch through the solenoid.
The HF is a quartz-fiber Cherenkov calorimeter with steel absorbers located outside the solenoid, on both sides of CMS, at about $\pm$11\m from the interaction point and extends the geometric coverage of the calorimeters to $\abseta<5.2$.
Unlike signals in the HB and HE, the signals in the HF are contained entirely within a 25\ns window. 
A detailed description of the CMS detector and the coordinate system used is presented in Ref.~\cite{Chatrchyan:2008aa}.

The HB has an approximately cylindrical structure that extends radially from $r=1.806$ to 2.950\m, and consists of 36 wedges covering the full azimuthal angle ($\phi$) range. 
The detector is divided into two cylindrical halves symmetrically about $z=0$; on the positive side is the HB plus, and on the negative side is the HB minus.  
Each wedge is made up of 14 copper alloy absorber plates and 2 stainless steel plates at the inside and outside faces, alternating with 17 layers of plastic scintillator tiles.
The thickness of the brass plates is 5\cm and that of the scintillating tiles is 3.7\mm, except the first tile which is 9.0\mm thick.
The HB readout has a symmetric 72-fold segmentation along the $\phi$ direction, and it is evenly segmented into 32 projective divisions in the $\eta$ direction (16 each for the HB plus and HB minus); 
thus, each projective unit in $\eta$-$\phi$ space (called a ``tower'') has a lateral dimension of $0.087{\times}0.087$, where $\phi$ is measured in radians.

The materials and structure of the HE are similar to those of the barrel system. 
There is one HE calorimeter on either side of the HB, denoted HE plus and HE minus. 
Each endcap consists of 18 wedges in the $\phi$ direction, and covers and closes one end of the barrel. 
The HE is constructed of plates, separated by staggered spacers, that are perpendicular to the beam axis.  
There are a total of 19 brass absorbing layers of width 8\cm in the HE, which provide as much as 9 interaction lengths of material for particles produced at the collision point. 
The projective towers in the HE have a lateral segmentation in $\eta$-$\phi$ space of $0.087{\times}0.087$ ($0.17{\times}0.17$) for $\abseta<1.6$ ($\abseta>1.6$).

\begin{figure}[!htp]
\centering
\includegraphics[width=0.8\linewidth]{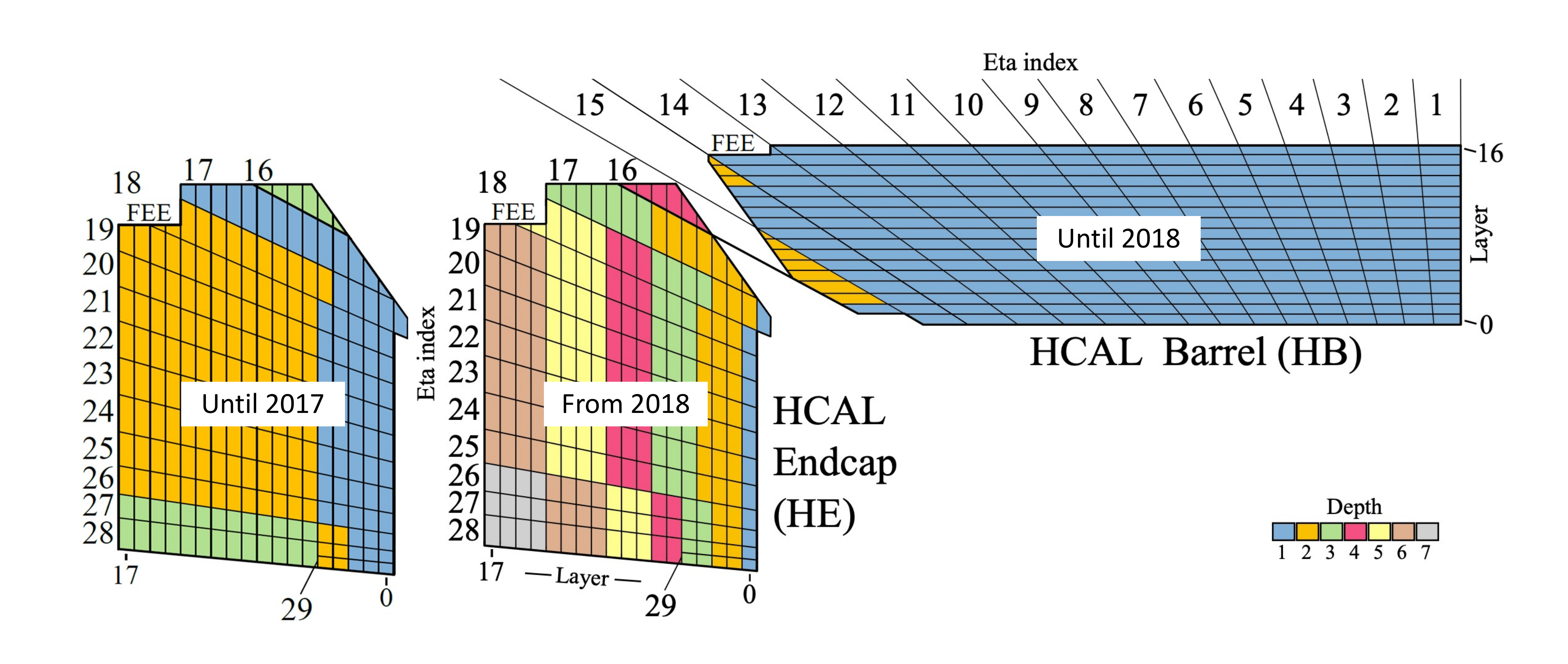}
\caption{A cross sectional view of the HB and HE in the $r$-$z$ plane. The \ieta coordinates are labeled, and the depth segmentation is coded in different colors.  The number of depths of the HE was increased before the 2018 data-taking period. The location of the front-end electronics (labeled ``FEE'') is also shown.}
\label{fig:HBHE_schematic}
\end{figure}

The readout of the HB and HE towers is subdivided radially into separate depths, each of which corresponds to a number of consecutive scintillator layers. 
The light produced in the plastic scintillating tiles from particles traversing that element of the detector is
collected in wavelength-shifting (WLS) fibers, optically summed, and sent to the photodetectors and front-end electronics where it is converted into a digital electric signal for data processing. 
From the perspective of data processing, a detector element, or ``channel,'' in the HB and HE can be uniquely 
identified by its location in $\eta$-$\phi$ space along with its depth. 
Integer indices for both $\eta$ and $\phi$ (\ieta and \iphi, respectively) are used to designate that location. 
The value of \iphi runs from 1 to 72, whereas \ieta runs from $+$1 to $+$29 or $-$1 to $-$29 in the plus and minus sides, respectively. 
The depth segmentation of the HE was modified during a year-end technical stop between 2017 and 2018. 
Figure~\ref{fig:HBHE_schematic} shows a cross-sectional view of the HB and HE in the $r$-$z$ plane, illustrating the depth segmentation in the HE before and after the 2017--2018 technical stop.

Hybrid photodiodes (HPDs) were used as the photodetectors in the HB and HE throughout Run 1, corresponding to the data-taking period from 2010 till 2012.
The HPDs were chosen for their magnetic field tolerance, high gain and linear response over a large dynamic range. 
However, they also exhibit high-amplitude anomalous noise from electrical discharges and ion feedback~\cite{HCALnoise}. 
As part of the HCAL Phase-1 upgrade~\cite{HCALtdrphase1upgrade}, the HPDs in the HE were fully replaced by silicon photomultipliers (SiPMs)~\cite{sipm} during the 2017--2018 technical stop (the HB was upgraded in a similar fashion during the long shutdown after 2018). 
The SiPMs provide a high gain between $10^4$ and $10^6$, have good quantum efficiency between 20 and 40\%, and operate well within a high magnetic field. 
The higher gain of SiPMs compared with HPDs, as well as their reduced size and power consumption, also allows for finer depth segmentation. 
In turn, increased depth segmentation improves shower energy resolution and helps mitigate the effects of radiation damage, since the most significant darkening of plastic scintillator occurs at high \abseta and depths closest to the collision point, and the lowered response in those affected channels can be calibrated out separately. 
Finally, the increased depth segmentation aids in the identification of in-time pileup, caused by additional $\Pp\Pp$ interactions distinct from the collision of interest and within the same bunch crossing, since such pileup tends to deposit its energy in the shallower depths.

The analog signal from the photodetectors in the HB and HE is digitized by a charge integrator and encoder (QIE) analog-to-digital converter (ADC) chip~\cite{Zimmerman2004TheDO}, which integrates the charge from the photodetector over a 25\ns interval.
This interval, which corresponds to the LHC bunch spacing, is called a ``time sample'' (TS).
Typically, HPDs will provide about 5\fC of input charge to the QIE for 1\GeV of energy deposit in a single HB/HE channel~\cite{CMSHCAL:2007zcq}, whereas SiPMs  provide about 1300\fC for the same energy~\cite{HCALtdrphase1upgrade}.
The charge is integrated by the QIE using a modified floating-point (pseudo-logarithmic) concept in which the input signal is divided into subranges with only a fraction of the signal being integrated for each subrange.
This allows the QIE to maintain a large effective dynamic range, while keeping the uncertainty from digitization subdominant to the energy resolution.
The internal timing of each channel is adjusted so that all channels observe a uniform arrival time within approximately 1\ns for particles originating at the nominal center of the channel.

During the 2017--2018 technical stop, the QIEs in the HE were also upgraded from a 7-bit encoding device (QIE8) to an 8-bit encoding device (QIE11); a 6-bit time-to-digital converter of the pulse arrival time was also included in the new QIE11.
A maximum relative quantization error of 2\% (1.4\%) for the QIE8 (QIE11) in the upper ranges of input signal is achieved.
Prior to 2018, 10 sequential TSs digitized by the QIEs were recorded in the data stream. 
The number was reduced to 8 TSs in 2018 in order to reduce the data volume, with minimal impact on the performance.
In both cases, the sample of interest (SOI), defined as the TS where the triggered event is placed, corresponded to the fourth TS in the window. 

\section{Local reconstruction algorithms}\label{sec:methods}
The main purpose of the HCAL local reconstruction algorithms is to estimate the energy deposited in a given channel in the SOI. 
Similar algorithms are employed by the ECAL~\cite{CMS:2020xlg}.
Understanding how the resulting pulse, as measured by the front-end electronics, is distributed as a function of time is critical. 
The intrinsic pulse shape in the HB and HE is affected by a number of factors, including the scintillation process in the tiles, the optical transmission in the WLS fibers, the photodetectors, and the QIE devices. 
Notably, the QIE introduces a ``time slew'' delay in the pulse shape that can be approximated as a logarithmic dependence on the total integrated charge $Q$ with respect to the start of the TS,
\begin{linenomath}
\begin{equation}
\Deltaslew = 11.98 - 1.56  \ln (Q),
\end{equation}
\end{linenomath}
where \Deltaslew and $Q$ are in the units of \ns and \fC, respectively, although the real amplifier slew rate actually depends on the (unknown) instantaneous input current. The value of \Deltaslew is constrained to be positive and less than 10\ns. The pulse shapes of SiPMs are different from HPDs, in part due to the higher gain.

\begin{figure}[htp]
\centering
\includegraphics[width=0.6\linewidth]{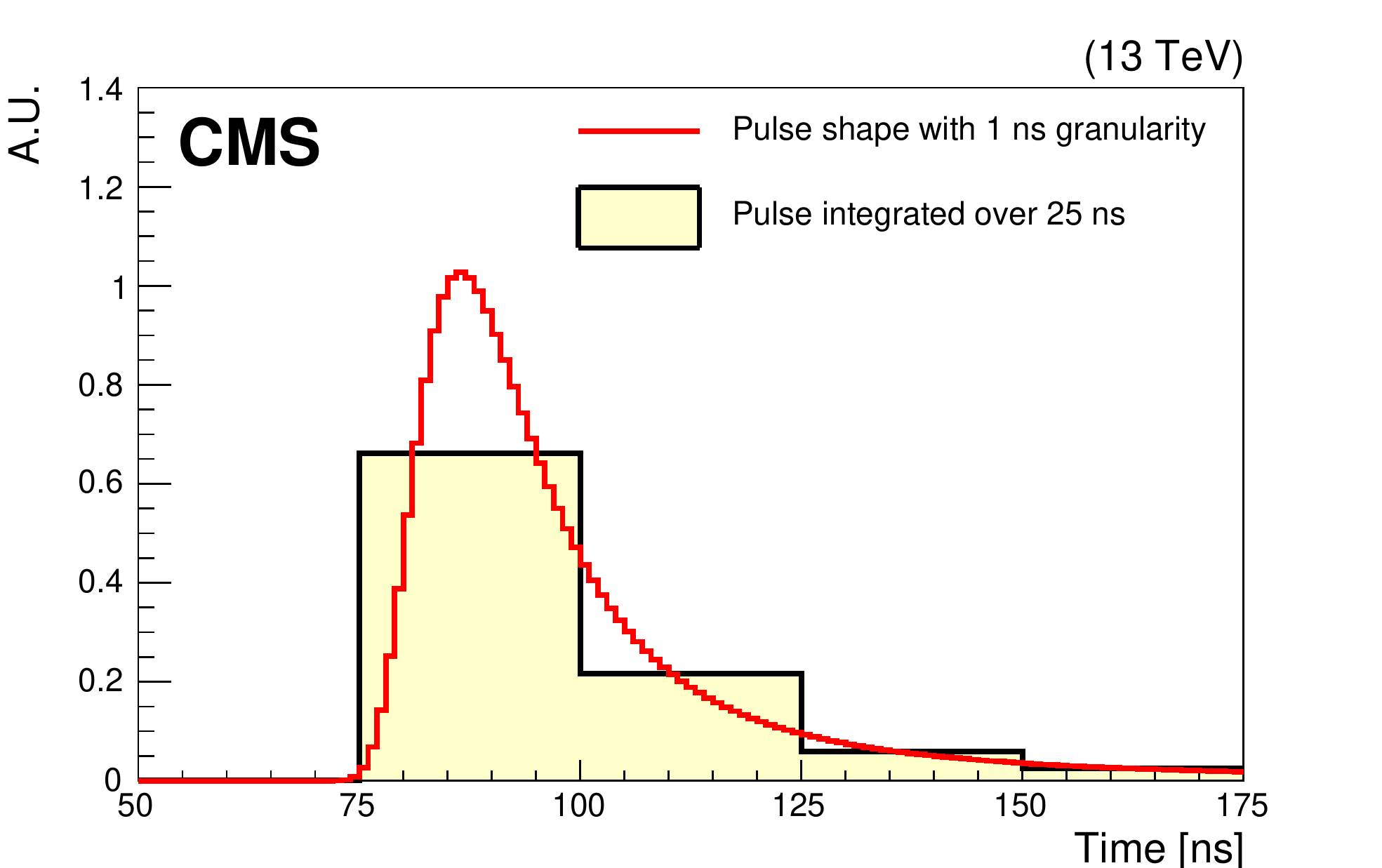}
\caption{
    Average pulse shape for high-energy depositions in the HE.
    The red solid line is the pulse shape used in reconstruction algorithms with 1\ns granularity.
    The yellow filled histogram is constructed from the red shape by integrating over each 25\ns TS. 
    The SOI corresponds to the TS from 75 to 100\ns. 
}
\label{fig:avgpulseshape}
\end{figure}

The pulse shape can be extracted with a 1\ns resolution from both test beam data and \textit{in situ} with $\Pp\Pp$ collision data. 
The pulse shape with HPDs was measured with a 300\GeV pion test beam prior to LHC collisions~\cite{Collaboration_2010}, while the pulse shape with SiPMs was extracted from pp collision data using isolated bunch collisions.
The extraction of the pulse shape was performed by adjusting the time settings of the QIE in 1\ns increments and measuring the pulses with different phases.
Figure~\ref{fig:avgpulseshape} shows the pulse shape as a function of time for high-energy depositions in the HE as measured by a SiPM, integrated over 1\ns and 25\ns bins. The pulse shape shown is an average over all channels in the HE.

\begin{figure}[htp]
\centering
\includegraphics[width=0.45\linewidth]{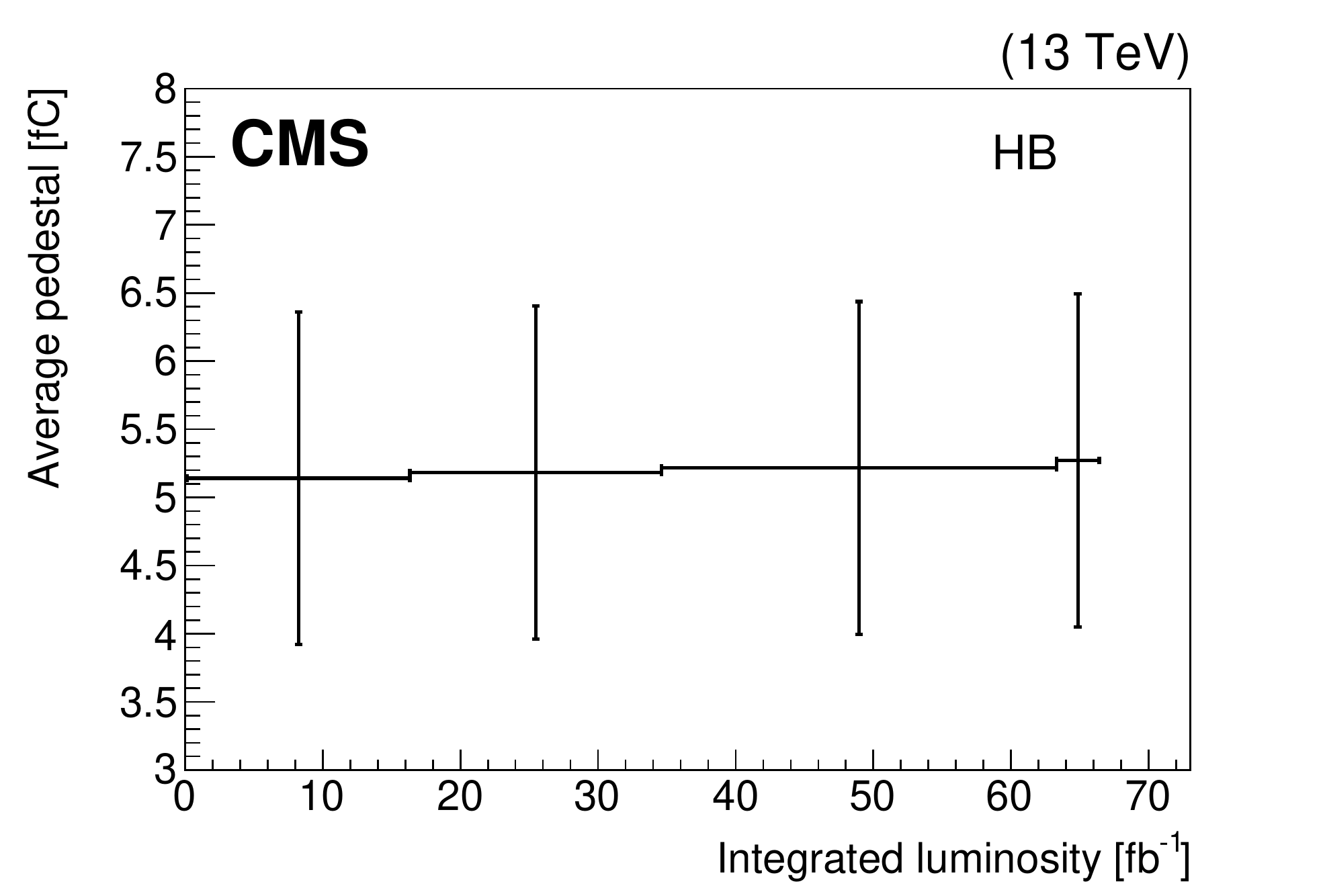}
\includegraphics[width=0.45\linewidth]{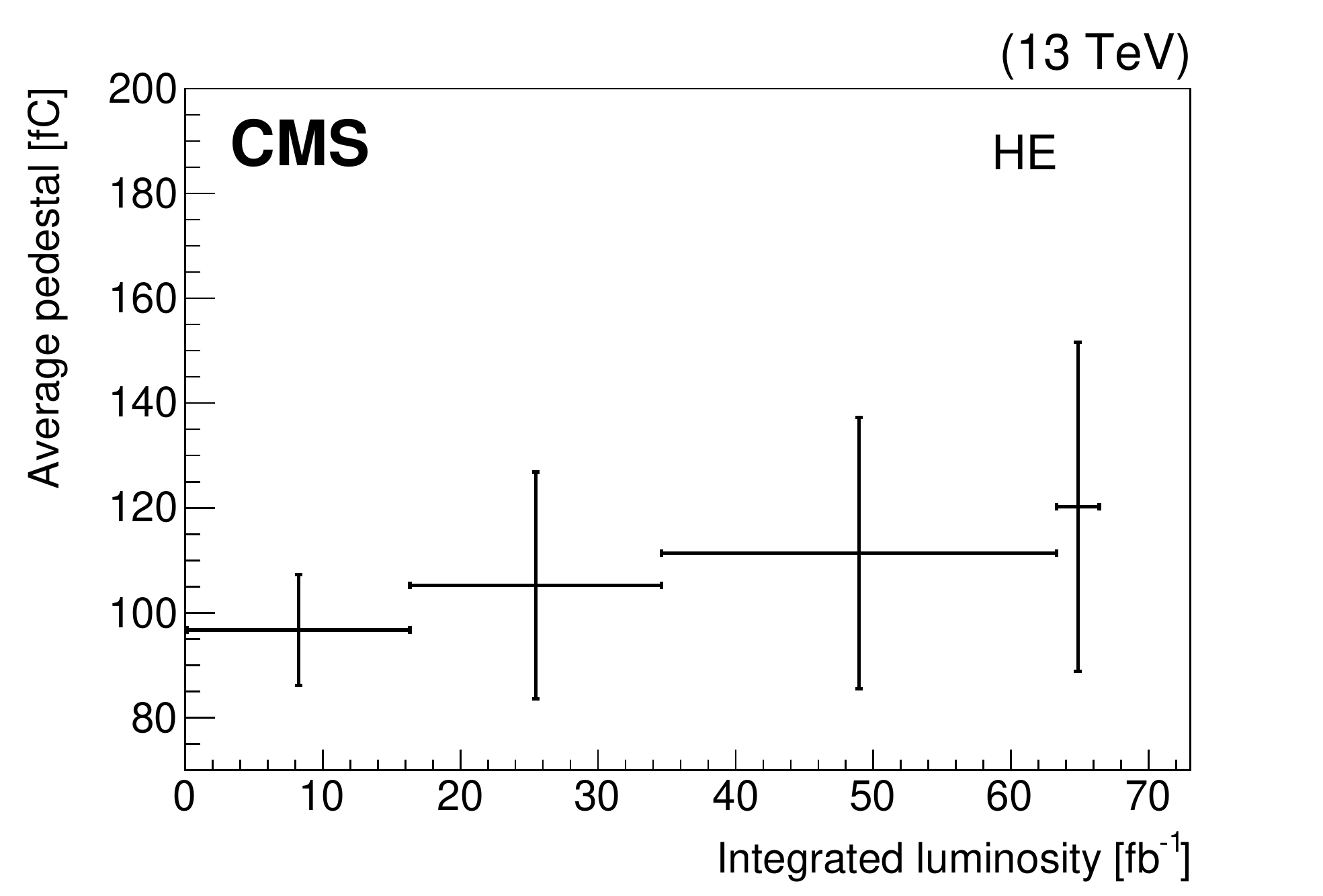}
\caption{
  The average value of the pedestal, as a function of the integrated luminosity, since the start of 2018 data taking for the HB (left) and HE (right). 
  Each horizontal line denotes the time range where the corresponding pedestal measurement is used in the energy reconstruction, and the error bars indicate the average standard deviation of the pedestal.
}
\label{fig:pedestal}
\end{figure}

Fluctuations in the pulse shape can be attributed to a variety of causes. 
The QIEs have a small nonzero energy reading even in the absence of a signal. 
Similarly, photodetectors exhibit increasing dark current on account of radiation damage. 
The sum of these two effects is referred to as the pedestal, whose average value and standard deviation are measured in dedicated pedestal runs, when beams are absent and where other subdetectors are not involved. The contribution of pickup noise to the QIEs from other subdetectors is negligible. 
Figure~\ref{fig:pedestal} shows the average value of the pedestal as a function of the integrated luminosity since the start of 2018 data taking for the HB and HE.
The pedestal values differ in the two detectors primarily because of the different photodetectors used.
Each horizontal line denotes the time range where the corresponding pedestal measurement is used in the energy reconstruction, and the vertical error bars indicate the average of the standard deviation of the pedestal distributions measured in individual channels.
The pedestals used in the energy reconstruction are correlated in time, being identical every four TSs due to the rotation of the four capacitors in each QIE.
At the start of 2018 data taking, the HE SiPMs had low dark current, so the pedestal values primarily come from the QIE, whose fluctuations are relatively small.  
However, by the end of 2018, the SiPM dark current introduced large fluctuations to the pedestal values.

The QIE also introduces a quantization uncertainty because of the finite resolution of its encoding. 
Finally, the finite number of photoelectrons collected introduces a Poisson uncertainty, which is approximated as $\sqrt{\smash[b]{\Npe}}$, 
since the number of photoelectrons is typically large ($\Npe \sim 32$ for a 1\GeV signal with SiPMs, whereas $\Npe \sim 6$ for a 1\GeV signal with HPDs).

The arrival time of the pulse shape can also vary for a variety of reasons, such as fluctuations in shower development. 
This has been measured in both test beam and collision data by fitting the appropriate template with a floating arrival time and amplitude (adjusted for the time slew) to individual channels and calculating the standard deviation of the distribution. 
The variation in the arrival time is larger at low energy; for energies $\gtrsim$5\GeV, the arrival times have a standard deviation of approximately 3\ns in the HE and 5\ns in the HB during the 2018 run. The HE has less variation than the HB because of the finer depth segmentation, which provides a more precise measurement of the shower development.
Long-lived particles can also result in a delayed arrival time leading to mismeasured energy.
Moreover, they may be especially sensitive to the OOTPU mitigation efforts that can differ between the trigger and offline reconstruction.
Simulation studies are used to estimate the size of this effect in beyond the standard model searches~\cite{CMS:2020iwv,Sirunyan:2678845,Sirunyan:2633728}.

\subsection{Method~0}
Prior to 2015, the LHC bunch-crossing spacing was kept at a minimum of 50\ns during normal data taking, so a simple sum of the charge in the SOI with 
the charge in its subsequent TS accurately reflects the energy initially deposited in the SOI.
To improve the precision, the average pedestal is subtracted from the total, and it is then corrected 
by a multiplicative factor to account for the 10--15\% of the energy outside of the two TSs. 
This algorithm, called Method~0 (M0), is unsuitable when
the bunch-crossing spacing is 25\ns because of potentially large additional contributions from the bunch crossings directly neighboring the SOI.

Three new algorithms were eventually deployed at CMS over the subsequent years, all using pulse-shape templates to extract the energy from the SOI.  
During 2016--2017, an algorithm referred to as Method~2 (M2) was used in the offline reconstruction, whereas Method~3 (M3) was used for online reconstruction in the HLT.  
Both algorithms were superseded in 2018 in both online and offline reconstructions by an algorithm called ``Minimization At HCAL, Iteratively'' (MAHI). An algorithm referred to as Method~1 was also developed, but because of its oversimplification, such as ignoring the time slew, it never performed well enough to be used.

\subsection{Method~2} 
The M2 algorithm estimates the energy in the SOI by minimizing a $\chi^2$ defined as
\begin{linenomath}
\begin{equation}
\label{eqn:method2_chi2}
  \chi^2 = \sum^1_{j=-1}\frac{(t_j - \tmean)^2}{\sigmat^2} + \frac{(\text{ped} - \pedmean)^2}{\sigmaped^2} + \sum^{\NTS-1}_{i=0}\frac{(A_i - \mimutped)^2}{\sigmapi^2}.
\end{equation} 
\end{linenomath}
Up to three separate fitted pulse shapes are used: one for the signal arriving in the SOI and one each for signals arriving in the adjacent TSs (\SOIm and \SOIp). 
The fit extracts the pulse amplitudes, $\vec{\mu}$, and the corresponding times of arrival, $\vec{t}$, as well as the pedestal, where the elements of the vector correspond to each of the three pulses.
The first term in the equation reflects a constrained fit for the arrival time of each pulse, $t_j$, which can be shifted from the expected mean, \tmean, with a penalty to the $\chi^2$ equal to the square of the difference divided by the standard deviation of the arrival time for that channel, \sigmat. 
The second term is a constrained fit for the pedestal, assumed to be a constant additive offset across all TSs; 
here \pedmean and \sigmaped are the expected mean and standard deviation of the pedestal for that channel, respectively.  
In the third term, $A_i$ is the amplitude of the QIE measurement (in \fC) of the $i^\text{th}$ TS;
\mimutped is the sum of the amplitudes of the fitted pulse shapes, $\vec{\mu}$, and pedestal in that TS; \NTS is the number of consecutive TSs recorded from the QIE, and \sigmapi is the combined uncertainty due to the pedestal, quantization error, and photostatistics, computed separately for each of the $i$ measurements.
The number of TSs, \NTS, was ten prior to 2018, and eight from 2018 onwards. 
In this paper, for the sake of consistency, eight TSs are used when comparing M2 and MAHI.
The $\chi^2$ is minimized with the MIGRAD algorithm as implemented in \textsc{Minuit}~\cite{James:1975dr}.

The M2 minimization is performed over two iterations.  
In the first iteration, the fit is performed with only a single pulse-shape template from the SOI. 
If contributions from OOTPU are small (typically when the signal pulse is large), fitting with one template gives a good result. 
However, if the $\chi^2$ from the first iteration is larger than 15 and the sum of pedestal-subtracted charges in the SOI and in \SOIp is less than 100\fC for HPDs or 25\,000\fC for SiPMs (both corresponding to an energy around 20\GeV), a second iteration is performed using three pulse shapes. 
In order to reduce the overall CPU time, the algorithm runs only if the sum of charges in the SOI and in \SOIp is greater than zero, and if the sum of all pedestal-subtracted charges is also greater than zero. 

\subsection{Method~3}  
The M3 algorithm was developed because M2 did not meet the computational timing requirement of the HLT. However, the increased pileup conditions in Run~2 would have resulted in significantly degraded performance in the HLT had M0 been kept as the default online reconstruction algorithm. Jet trigger rates, for instance, would have increased manyfold without the development of M3.

One simplifying technique of the method is to assume a fixed arrival time for the pulses (after corrections for the time slew); another is to restrict the QIE measurements under consideration to three TSs: \SOIm, SOI, and \SOIp. 
Stochastic uncertainties from various sources are also ignored.
The problem can then be reduced from an iterative minimization algorithm to solving a system of linear equations: 
\begin{linenomath}
\begin{equation}
   \begin{bmatrix} 
   \ASOIm \\
   \ASOI  \\
   \ASOIp \\
   \end{bmatrix} 
   =
   \begin{bmatrix} 
   f_0(\ASOIm) & 0 & 0  \\
   f_1(\ASOIm) & f_0(\ASOI) & 0  \\
   f_2(\ASOIm) & f_1(\ASOI) & f_0(\ASOIp)  \\ 
   \end{bmatrix} 
   \begin{bmatrix} 
     \mu_{\SOIm} \\
     \mu_{\SOI}  \\
     \mu_{\SOIp} \\
   \end{bmatrix} 
   + 
   \begin{bmatrix} 
   B   \\
   B   \\
   B   \\
   \end{bmatrix},
\end{equation}
\end{linenomath}
where $A_i$ and $\mu_i$ are the QIE measurements after pedestal subtraction and the amplitude of the pulse for the indexed TS, respectively.
The functions $f_0$, $f_1$, and $f_2$ are the premeasured fractions of the pulse template contained in $+$0, $+$1, and $+$2 TS, respectively, and are functions of the amplitude of the pulse in the relevant SOI to correct for the corresponding time slew.
The baseline, $B$, is the average of the QIE measurements after pedestal subtraction in all TSs excluding SOI and \SOIp. 
The value of $B$ is not allowed to exceed three times the value of \sigmaped to avoid bias due to large early or late pulses.  

\begin{figure}[!htp]
\centering
\includegraphics[width=0.6\linewidth]{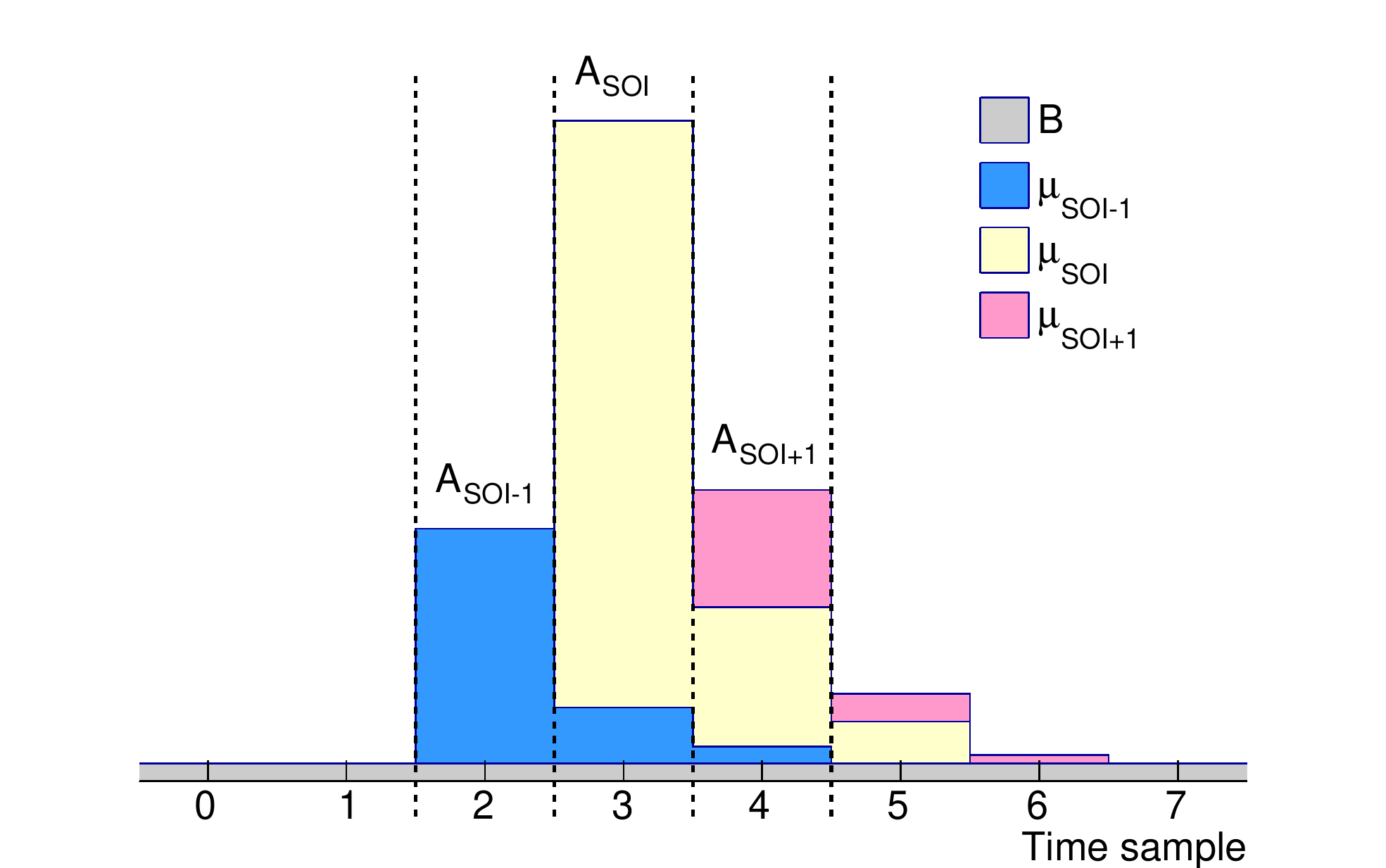}
\caption{
    An illustration of the M3 algorithm.
    Only three TSs from \SOIm to \SOIp, indicated with dash lines, are used in the reconstruction.
    Pulse-shape templates for \SOIm, SOI, and \SOIp are shown in blue, yellow, and pink, respectively.
    The gray shading shows the baseline.}
\label{fig:method3}
\end{figure}

Figure~\ref{fig:method3} shows an illustration of how the algorithm works to extract the pulse shape in the SOI.

\subsection{Minimization At HCAL, Iteratively}  
Although M2 and M3 use the same pulse-shape templates, the two algorithms are otherwise quite different, 
and the inconsistency between using M3 for online reconstruction and M2 for offline reconstruction introduces difficulties.
For instance, threshold effects for jet-based triggers are worsened, and electron and photon identification algorithms, which require a low value of the ratio between hadronic and electromagnetic energies, become less reliable. 
The MAHI algorithm provides precise measurements of the HCAL energy in the presence of OOTPU, which are suitable for offline reconstruction while still meeting the timing demands of the HLT.  

The MAHI algorithm constructs an $8{\times}8$ covariance matrix (corresponding to 8 TSs for each dimension) out of terms for the pulse-shape uncertainty (\Dpulse) and the noise (\Dnoise).
The noise term includes uncertainties due to QIE quantization, pedestals, and photostatistics. 
These uncertainties are added to the diagonal elements of the matrix, although correlations (and hence off-diagonal elements) will play an increasingly important role in Run~3, when the SiPM dark current is expected to increase.
For each pulse template, one covariance matrix \Dpulsej is constructed, leading to a total of eight covariance matrices.  
The final covariance matrix $\mathbf{V}$ is constructed according to
\begin{linenomath}
\begin{equation}
\mathbf{V} = \sum_{j=0}^{7} \mu_j^2 \Dpulsej + \Dnoise,
\end{equation}
\end{linenomath}
where $\mu_j$ is the amplitude of the pulse arriving in $\text{TS}_j$.

Then, a non-negative least squares algorithm~\cite{DBLP:0408029} is run to find $\mu_j$, whose values are constrained to be positive, by minimizing
\begin{linenomath}
\begin{equation}
    \chi^2 =
    \left[ \sum_j {\vec{P}_j \mu_j} - \vec{d} \right]^T 
    \mathbf{V}^{-1}
    \left[ \sum_j {\vec{P}_j \mu_j} - \vec{d} \right],
\end{equation}
\end{linenomath}
where $\vec{P}_j$ are the 8-element vectors that contain the contributions of the pulse templates to each TS, and $\vec{d}$ is the vector that contains the QIE measurements after pedestal subtraction. 
At the beginning, the covariance matrix is initialized with only the noise terms. 
After the first iteration, the covariance matrix is updated using the $\mu_j$ values that minimize the $\chi^2$ value, and the next iteration begins.
If the change in $\chi^2$ between two iterations is less than $10^{-3}$ or the number of iterations goes beyond 500, the iteration stops. 
Typically, the number of iterations is less than 10.
Thus, MAHI incorporates the information used in M2 and extends the number of pulse shapes under consideration (from 3 to 8) while still being able to run on the HLT within the time budget. 
For a typical event in the Run~2 data set with large hadronic activity, MAHI is $\mathcal{O}(10)$ times faster than M2, but still $\mathcal{O}(10)$ times slower than M3, independent of pileup.

\begin{figure}[!htp]
\centering
\includegraphics[width=0.45\linewidth]{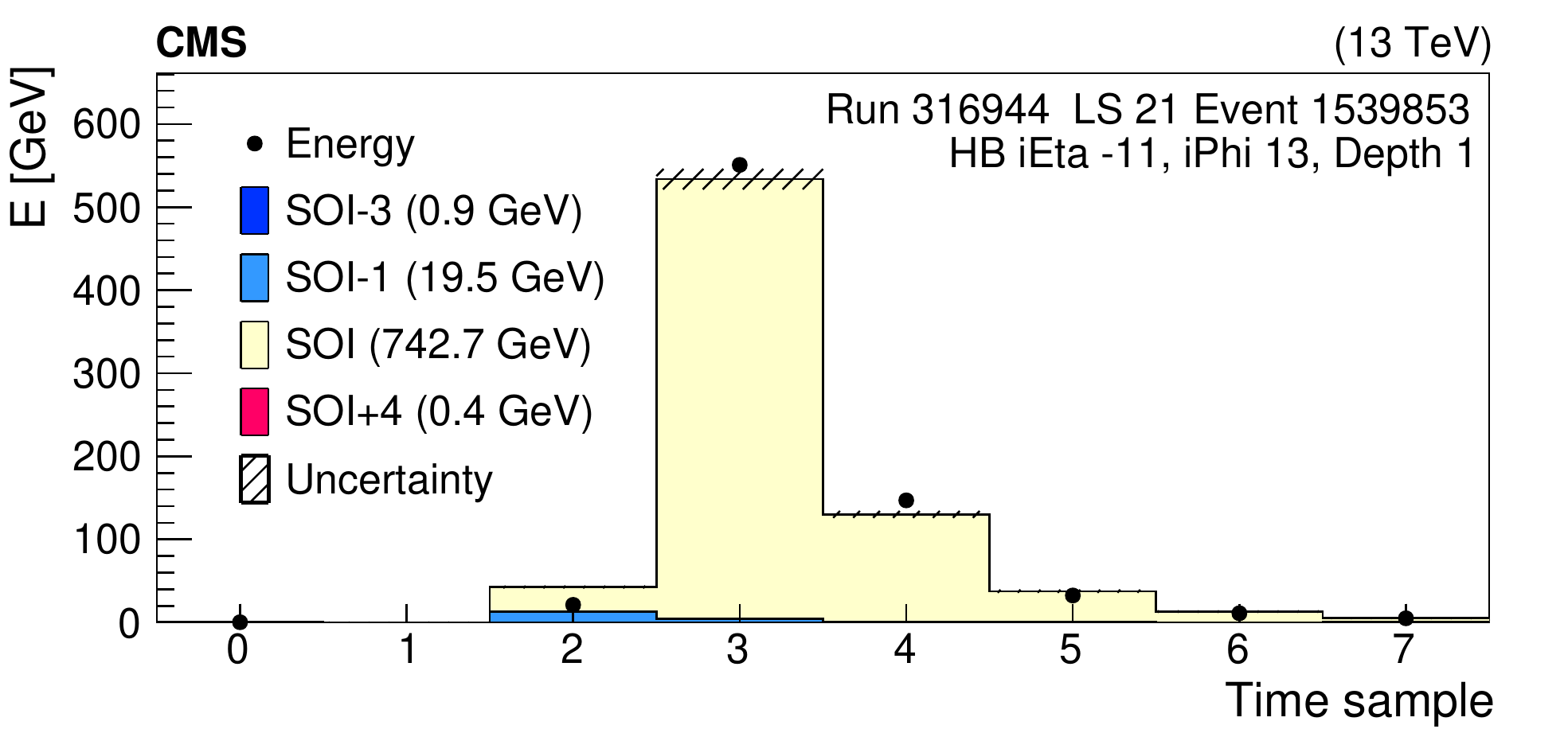}
\includegraphics[width=0.45\linewidth]{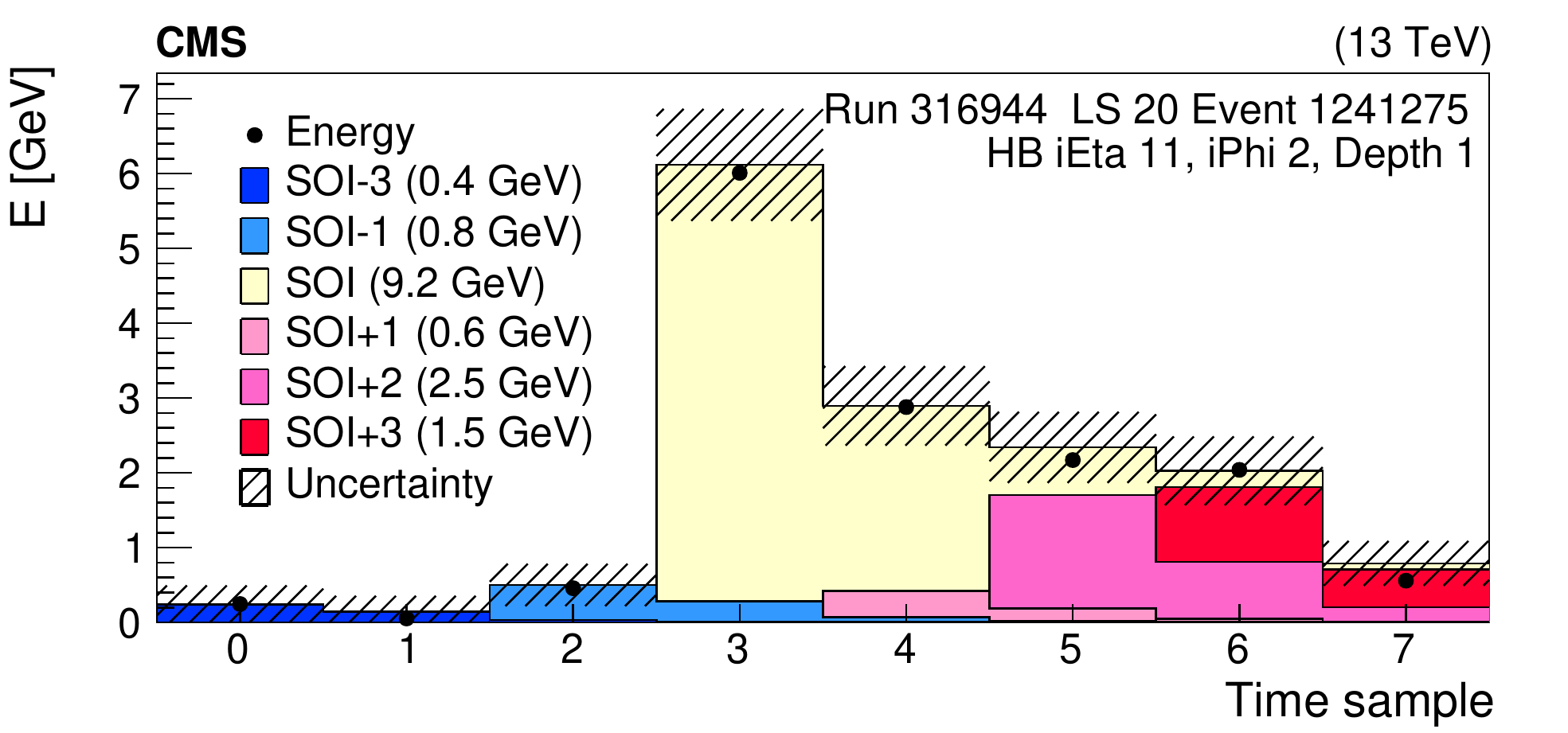}
\includegraphics[width=0.45\linewidth]{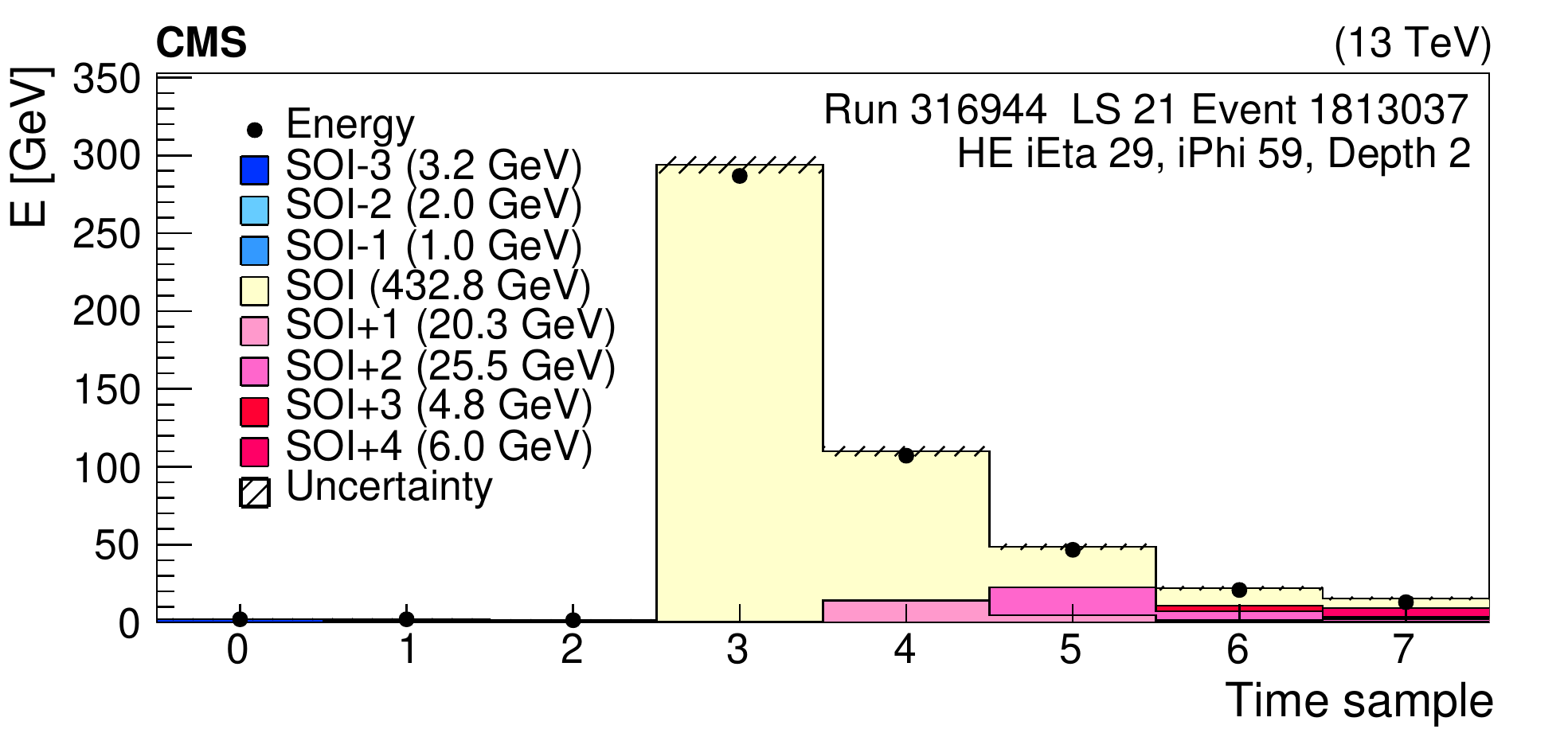}
\includegraphics[width=0.45\linewidth]{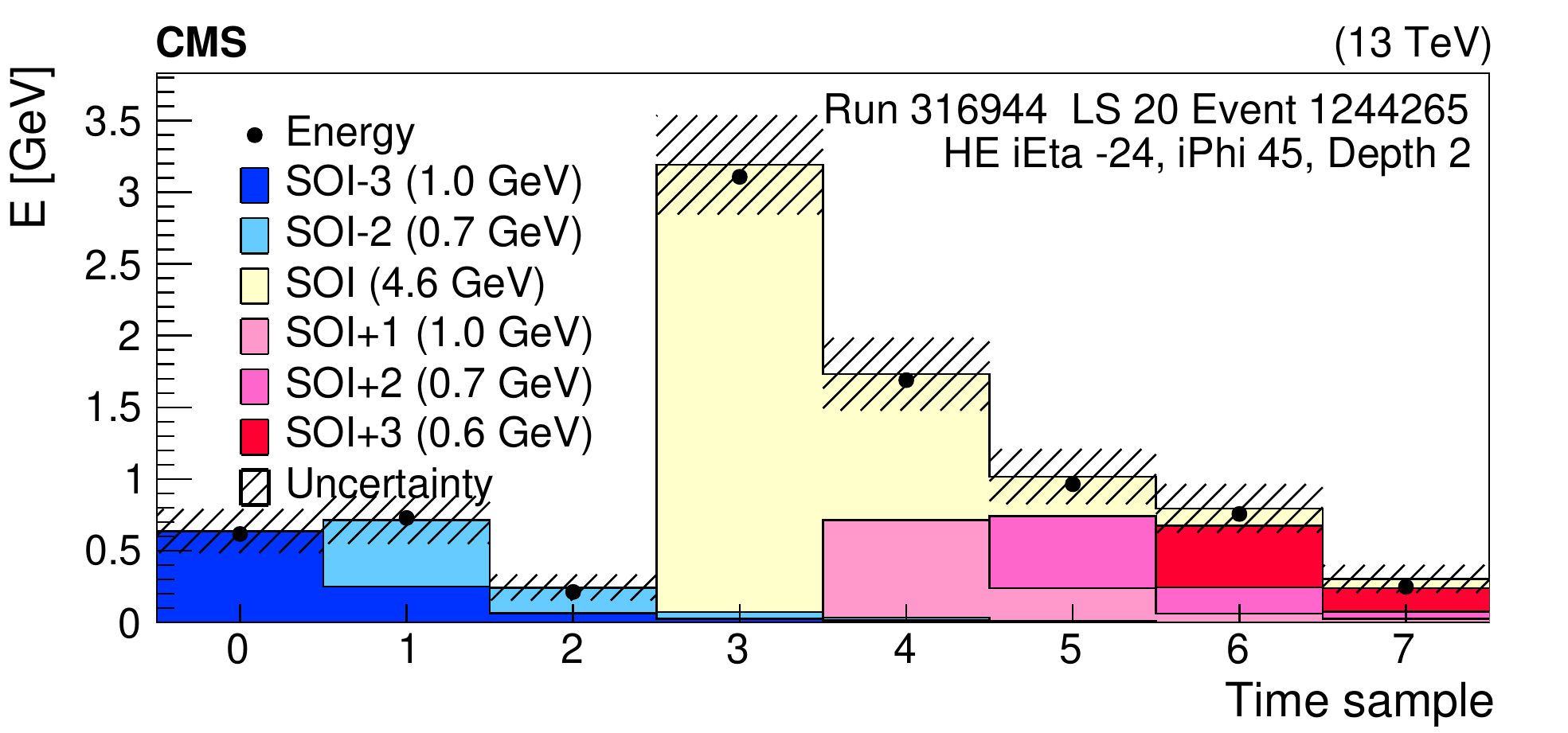}
\caption{
    Representative fit results from MAHI for high-energy (left) and low-energy (right) pulses, in the HB (upper) and the HE (lower). 
    The recorded pedestal-subtracted charge in the QIE (converted to units of energy) is given by the points, while the filled histograms represent the fitted values for the various pulse shapes. 
    The sum of the fitted energy for each pulse, labeled by its position relative to the SOI, is presented in the legend. 
    The combined uncertainty from the pedestal, photostatistics, and QIE quantization is shown by the hatched areas.}
\label{fig:pulseshape1}
\end{figure}

The fit results of MAHI using 2018 pp collision data with $\approx$50 average interactions per proton bunch crossing are illustrated in Fig.~\ref{fig:pulseshape1}. 
Representative fits at high and low energy and for both the HB (with HPD photodetector and QIE8 ADC) and HE (with SiPM photodetector and QIE11 ADC) are shown. 
The uncertainty band includes the QIE and SiPM leakage currents, photostatistics, and QIE quantization. 
When the energy deposition from the SOI is high and dominates over the other TSs, a single pulse shape provides a good fit; 
however, at lower energies, contributions from OOTPU are important and must be subtracted to provide a good estimate of the energy in the SOI.

\section{Reconstruction performance}\label{sec:performance}
\begin{figure}[!htp]
\centering
\includegraphics[width=0.45\linewidth]{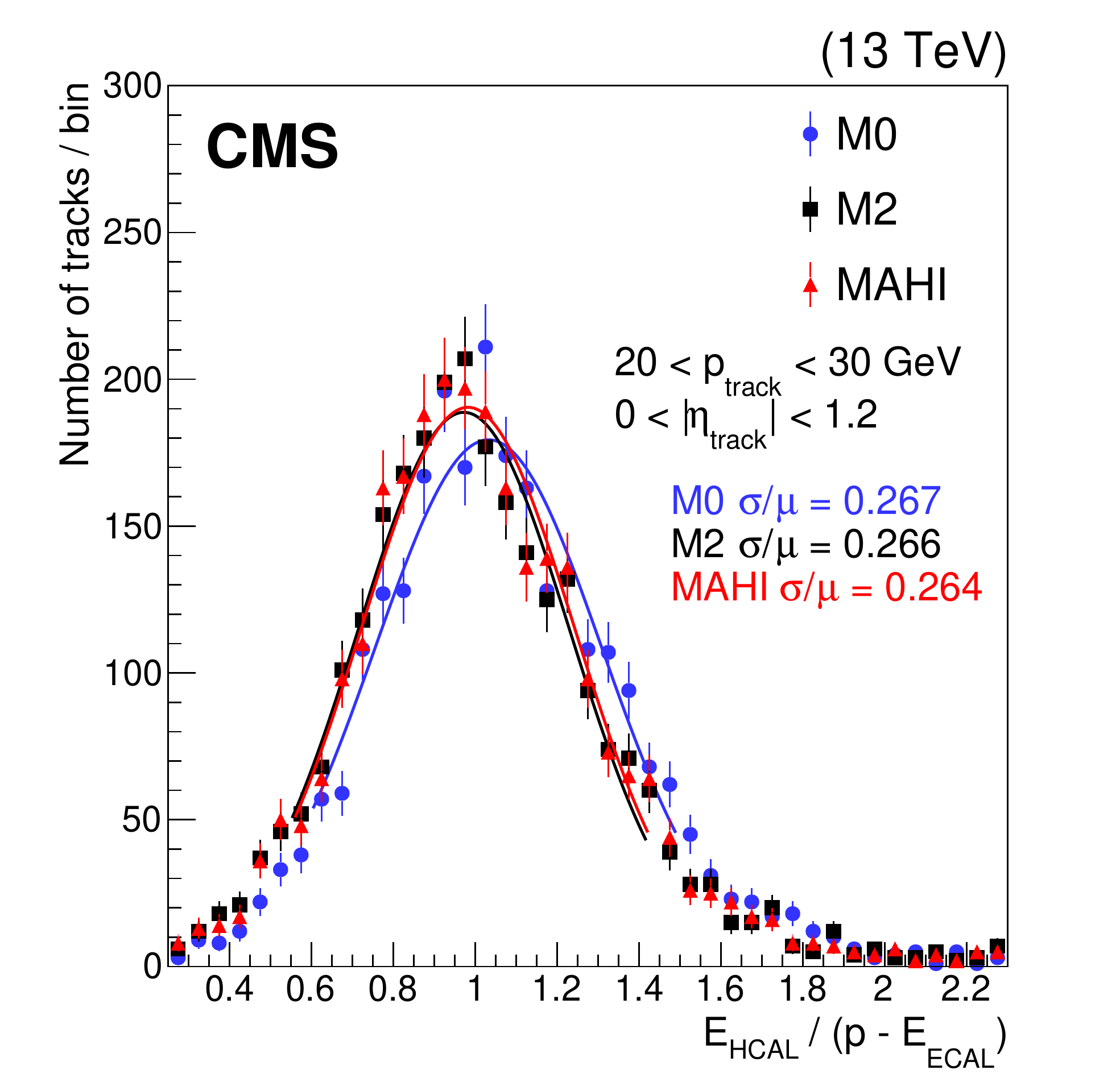}
\includegraphics[width=0.45\linewidth]{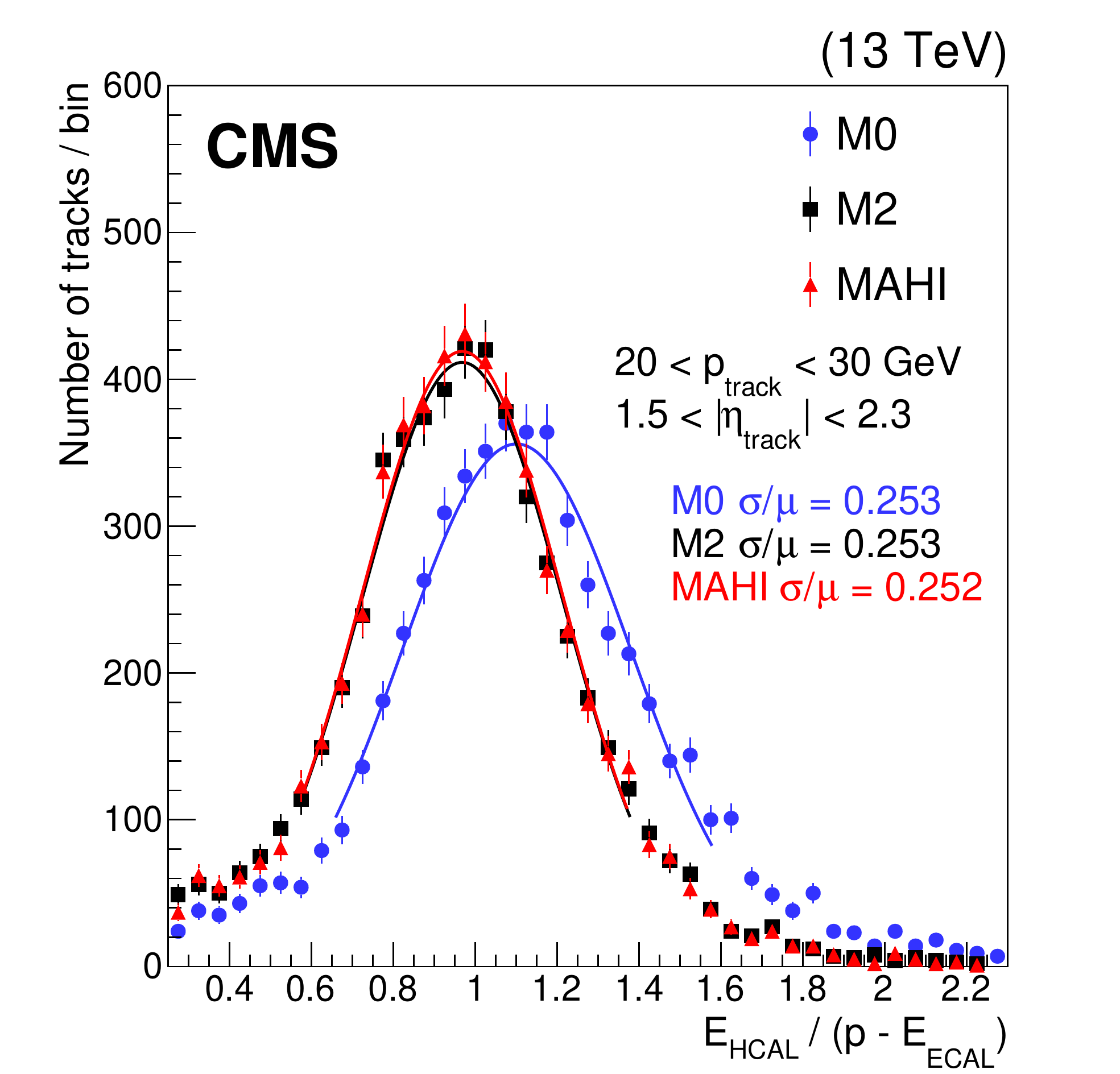}
  \caption{The HCAL energy response of M0, M2, and MAHI, measured in an electron/photon-triggered data set using isolated tracks with $20 < p_\text{track} < 30\GeV$ and either $\absetatrack<1.2$ (left) or $1.5 <\absetatrack<2.3$ (right). The vertical bars show the statistical uncertainty in the number of tracks in each bin. The measured energy resolution for each method in this sample is comparable.}
\label{fig:MAHI_vs_M0_20-30}
\end{figure}

The performance of the various algorithms used to reconstruct HCAL energy can be evaluated in a number of ways.
The removal of the OOTPU contribution has more significant effects at lower \pt and at higher \abseta.
This can be seen locally in the reconstruction of isolated charged hadrons, similar to the study in Ref.~\cite{CMS:2019hpr}.
Isolated tracks with momenta between 20 and 30\GeV are selected from a sample of events triggered by an electron or photon to avoid bias from the trigger.
These tracks are extrapolated through the calorimeters, and the energy depositions in calorimeter channels within a radius of 35\cm are clustered, typically capturing more than 99\% of the shower energy.
The associated ECAL energies are required to be less than 2\GeV. 
Figure~\ref{fig:MAHI_vs_M0_20-30} shows the ratio of the clustered energy in the HCAL to the track momentum minus the clustered energy in the ECAL for the various algorithms. The solid lines each represent a Gaussian fit to the core of the distributions. The fits are dominated by the HCAL resolution; the standard deviation is comparable for M2 and MAHI, which both subtract the OOTPU. At large $\eta$, the M0 response is higher due to OOTPU contributions; moreover, the energy response of M0 exhibits more prominent, non-Gaussian tails.

\begin{figure}[!htp]
\centering
\includegraphics[width=0.32\linewidth]{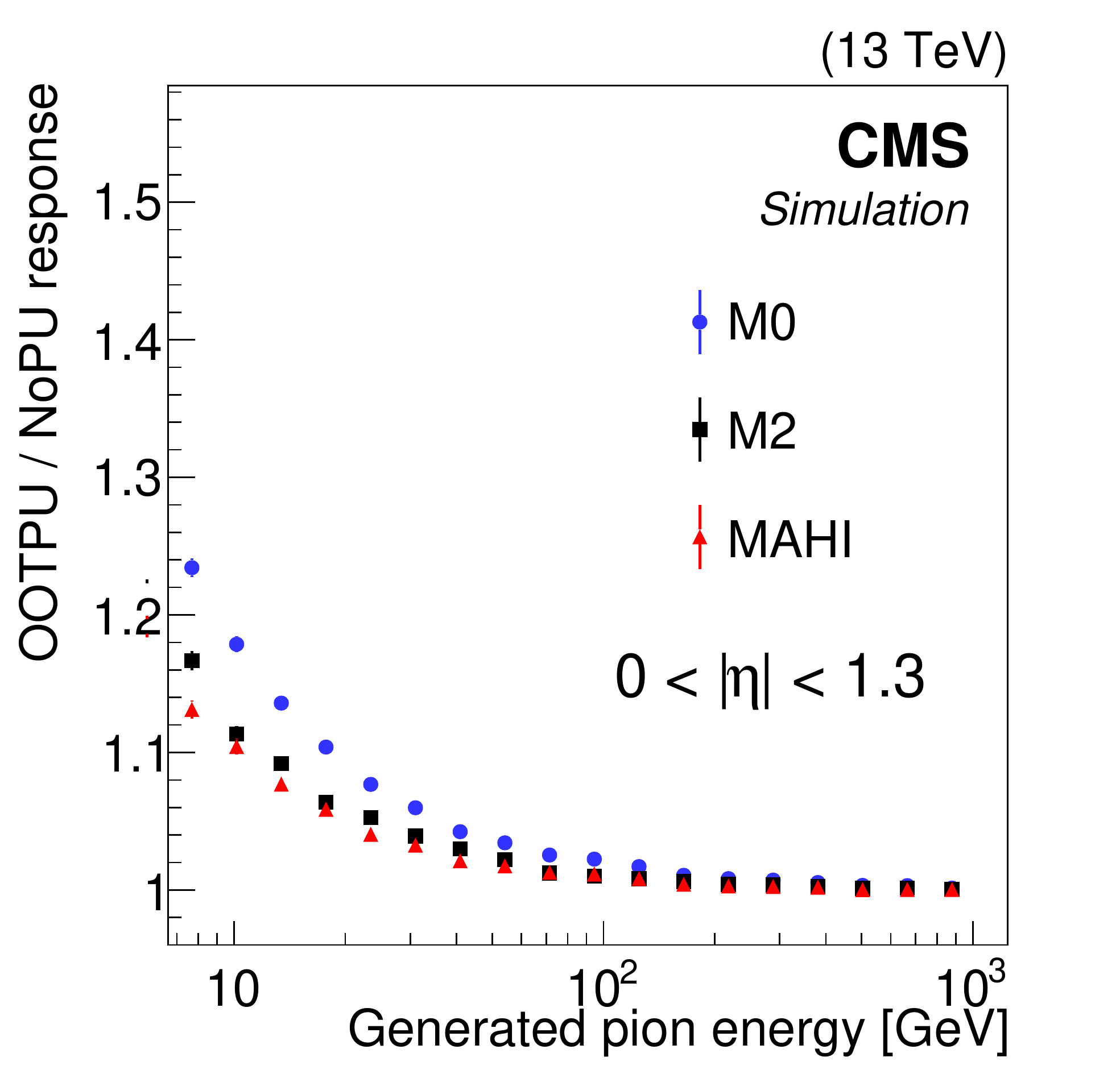}
\includegraphics[width=0.32\linewidth]{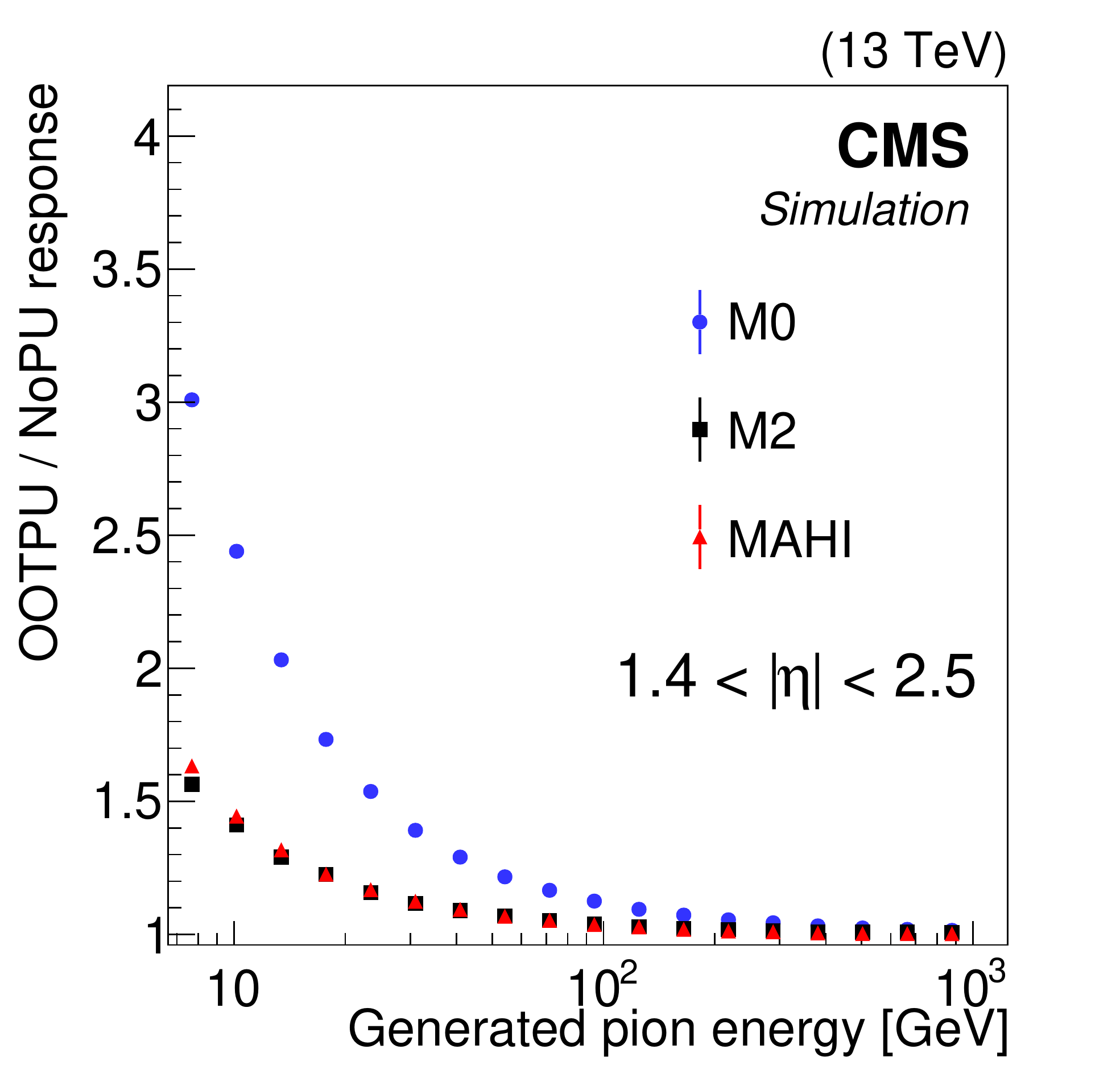}
\includegraphics[width=0.32\linewidth]{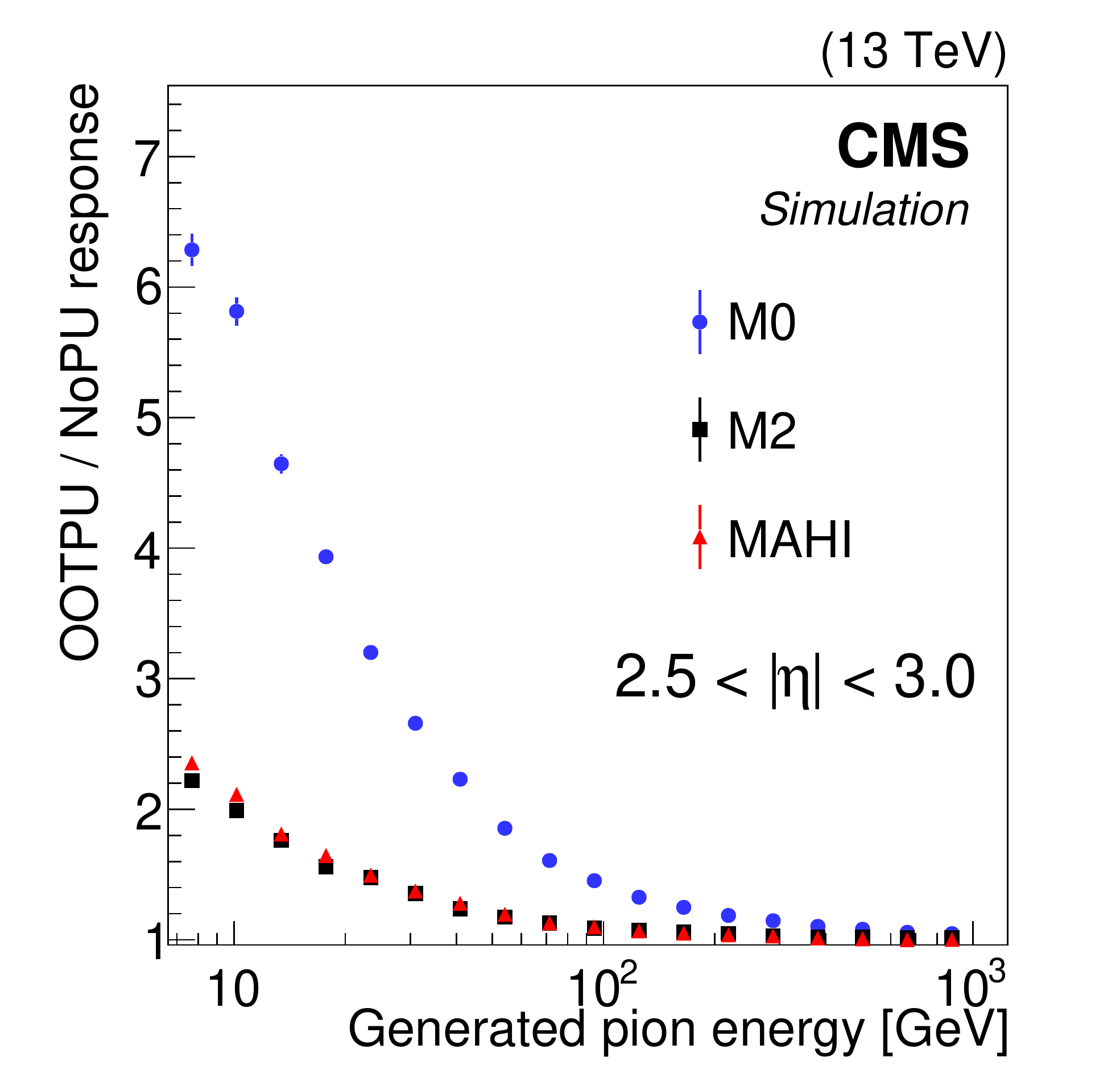}
\caption{Ratios of responses to simulated charged pions between a sample with OOTPU included to a sample with no pileup, as functions of the generated pion energy.
The left plot shows the response for pions in the HB ($\abseta < 1.3$), whereas the middle and right plots show the response for pions in the HE ($1.4 < \abseta < 2.5$ and $2.5 < \abseta < 3.0$).}
\label{fig:response_OOTPU_NoPU_ratio}
\end{figure}

The impact of OOTPU on the mean response is assessed directly in Monte Carlo simulation.
Single pions are generated with \PYTHIA version 2.242~\cite{pythia} at the interaction point with various values of momenta and pseudorapidity. The interaction of particles with the CMS detector is modeled using the \GEANTfour toolkit~\cite{geant4}.
Reconstructed energy deposits are clustered in a cone of radius $\Delta R=\sqrt{\smash[b]{(\Delta\eta)^2 + (\Delta\phi)^2}}=0.2$ about the generated pion's trajectory, and the clustered energy is divided by the generated energy to determine the response. 
Charged pions with clustered energy in the ECAL exceeding 1\GeV are excluded. 
The ratios of the response in a sample with OOTPU to a sample without any pileup is shown as a function of pion energy in Fig.~\ref{fig:response_OOTPU_NoPU_ratio}. 
The OOTPU generated in this sample corresponds to approximately 30 interactions per bunch crossing, typical of 2018 data, but there is no in-time pileup contribution simulated. 
The M0 algorithm exhibits a larger bias in its response at lower energies, which is worse at larger values of \abseta.
Neither M2 nor MAHI is able to achieve equal response at low energies because both algorithms are designed not to provide negative energies, which biases the average response in the positive direction.

\begin{figure}[!htp]
\centering
\includegraphics[width=0.45\linewidth]{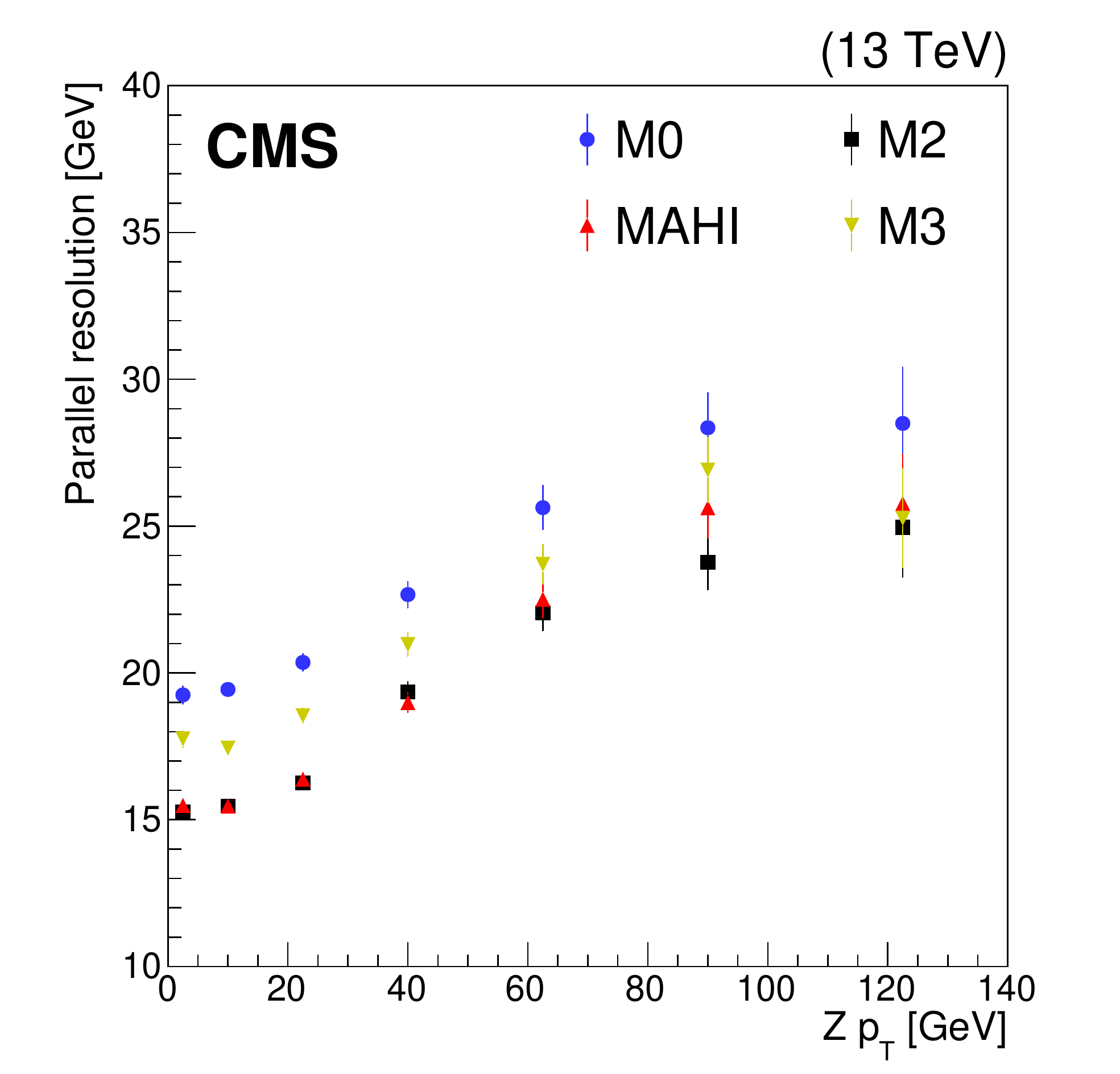}
\includegraphics[width=0.45\linewidth]{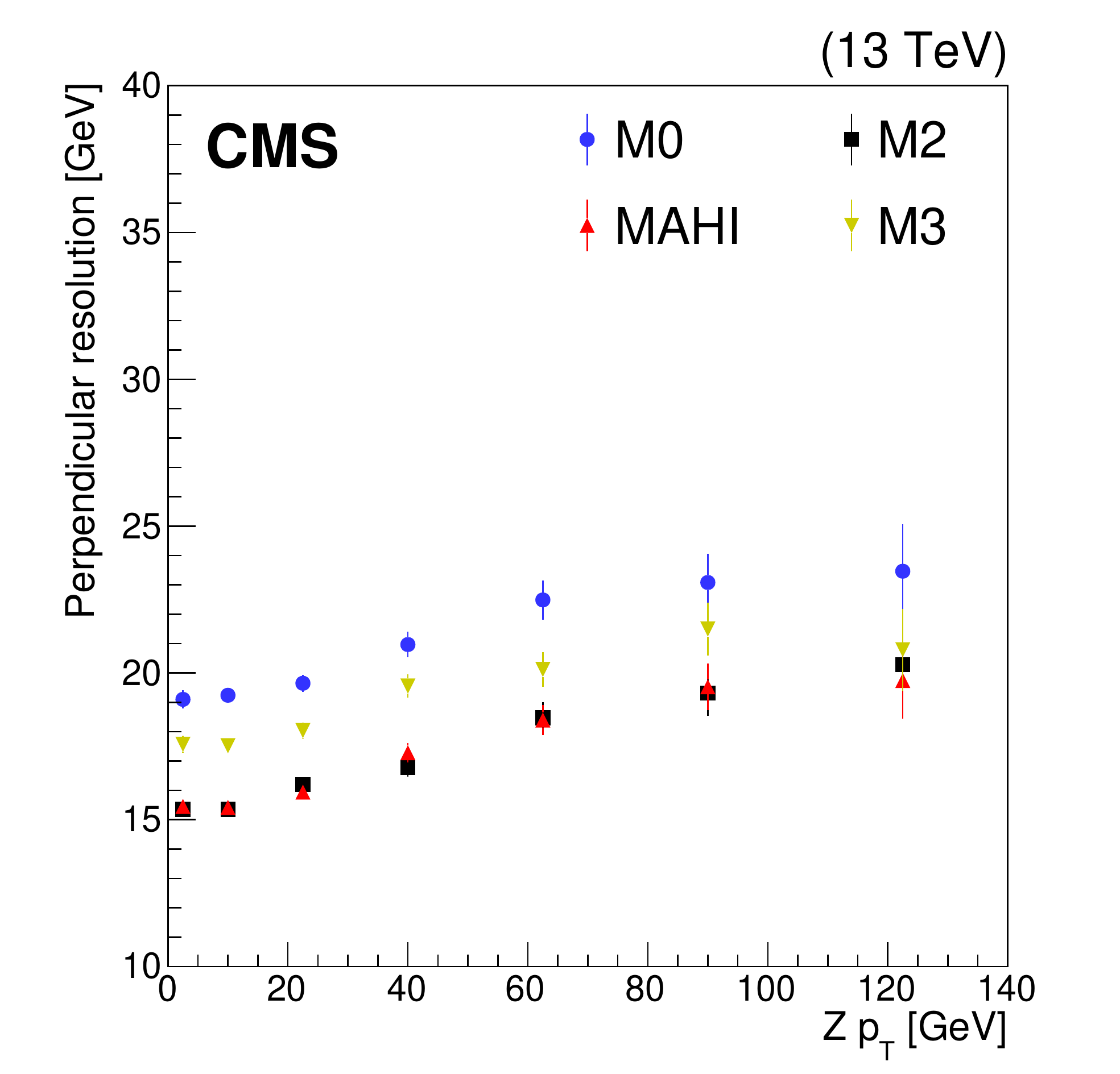}
\caption{\ptmiss resolutions as a function of the \PZ boson \pt, measured in a data set triggered by two muons.
The left (right) plot is the resolution of the component parallel (perpendicular) to the \PZ boson's \pt. Error bars reflect statistical uncertainties.}
\label{fig:CaloMETBE_data}
\end{figure}

The global performance of the HCAL energy reconstruction algorithms is evaluated in events containing a \PZ boson.
Such events have very little intrinsic \ptmiss, hence any reconstructed \ptmiss can be attributed primarily to the detector resolution.
The events are required to contain two isolated, oppositely charged muons with $\pt>20$ and 10\GeV, respectively, with their reconstructed invariant mass satisfying $81<M_{\Pgm\Pgm}<101\GeV$. 
The \ptmiss is calculated as the negative vector sum of the energies in the individual calorimeter towers of the ECAL and HCAL (excluding HF and HO). 
The parallel and perpendicular components of the \ptmiss are computed as projections with respect to the \PZ boson $\vec{\pt}$ direction, similarly to Ref.~\cite{MET_performance_paper}. 
Figure~\ref{fig:CaloMETBE_data} shows a comparison of the resolution of the parallel and perpendicular components of the recoil system between M0, M2, M3, and MAHI in 2018 data, with an average pileup around 30. 
The M2, M3, and MAHI algorithms demonstrate improved resolution in both components because of their ability to suppress OOTPU. 
Among these three algorithms, the marginally worse resolution of M3 was a motivation for switching to MAHI in 2018.

\begin{figure}[!htp]
\centering
\includegraphics[width=0.7\linewidth]{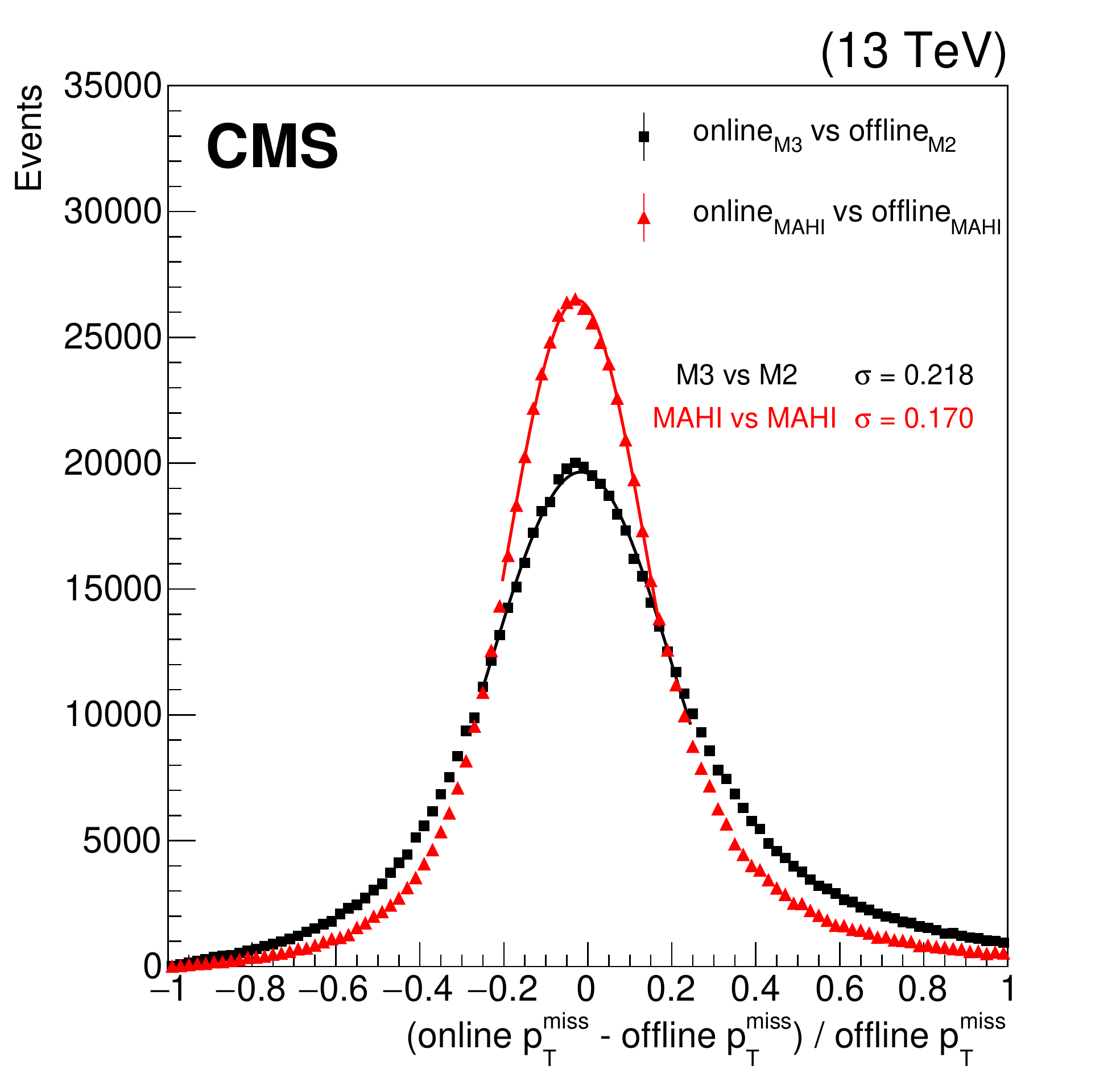}
\caption{Relative difference of \ptmiss in muon-triggered events between the online and offline reconstructions.
  Black: $(\text{online }\ptmiss(\text{M3}) - \text{offline }\ptmiss(\text{M2})) / \text{offline }\ptmiss(\text{M2})$. Red: $(\text{online }\ptmiss(\text{MAHI}) - \text{offline }\ptmiss(\text{MAHI})) / \text{offline }\ptmiss(\text{MAHI})$. The differences are fit to a Gaussian distribution and the fitted value of the standard deviation is shown.}
\label{fig:mahi-m2m3}
\end{figure}

To demonstrate the improvements from using consistent local reconstruction algorithms online and offline, Fig.~\ref{fig:mahi-m2m3} shows the relative difference of \ptmiss in muon-triggered events between the online and offline reconstructions, along with fits with a Gaussian function and the corresponding $\sigma$.
The \ptmiss is calculated as the negative vector sum of the energies in the ECAL and HCAL (this time including HF, but still excluding HO) and the muon momenta.
The average \ptmiss in these events is approximately 40\GeV. 
In one case, M3 and M2 are used online and offline, respectively, whereas in the other case, MAHI is used for both. 
The agreement between online and offline \ptmiss improves significantly when using MAHI consistently at both levels. 
The residual online-offline difference observed with MAHI is mostly attributable to the calorimeter calibrations, which, for both HCAL and ECAL, differ because the calibrations used in the HLT cannot be updated retroactively. 
There are also slight differences in the configurations of the MAHI algorithm.
This same inconsistency of calibrations applies to the combination of M3 online and M2 offline as well.
The additional penalty from using inconsistent algorithms is significant and further increases the differences between online and offline reconstructions, which can result in less efficient event triggering.

\section{Summary}\label{sec:summary}
Four local energy reconstruction algorithms for the CMS hadron calorimeter (HCAL) have been presented in this paper and their performance compared. 
When the bunch-crossing spacing is at least 50\ns, Method~0 performs well; however, a pulse-shape fitting algorithm must be used when the spacing is only 25\ns. 
The performance of Method~2 is strong under conditions of high out-of-time pileup, but its long reconstruction time makes it unusable in the high-level trigger. 
The use of a different algorithm, such as Method~3, to accommodate the time constraints of online reconstruction is possible, but the deleterious effects of mismatched algorithms make this solution undesirable. 
``Minimization at HCAL, Iteratively'' is a pulse-shape fitting algorithm that readily suppresses out-of-time pileup and has good intrinsic energy resolution, and is also sufficiently fast to run in the high-level trigger. Hence, it was the preferred local energy reconstruction algorithm for HCAL by the end of Run 2.

\begin{acknowledgments}
  We congratulate our colleagues in the CERN accelerator departments for the excellent performance of the LHC and thank the technical and administrative staffs at CERN and at other CMS institutes for their contributions to the success of the CMS effort. In addition, we gratefully acknowledge the computing centers and personnel of the Worldwide LHC Computing Grid and other centers for delivering so effectively the computing infrastructure essential to our analyses. Finally, we acknowledge the enduring support for the construction and operation of the LHC, the CMS detector, and the supporting computing infrastructure provided by the following funding agencies: BMBWF and FWF (Austria); FNRS and FWO (Belgium); CNPq, CAPES, FAPERJ, FAPERGS, and FAPESP (Brazil); MES and BNSF (Bulgaria); CERN; CAS, MoST, and NSFC (China); MINCIENCIAS (Colombia); MSES and CSF (Croatia); RIF (Cyprus); SENESCYT (Ecuador); MoER, ERC PUT and ERDF (Estonia); Academy of Finland, MEC, and HIP (Finland); CEA and CNRS/IN2P3 (France); BMBF, DFG, and HGF (Germany); GSRI (Greece); NKFIH (Hungary); DAE and DST (India); IPM (Iran); SFI (Ireland); INFN (Italy); MSIP and NRF (Republic of Korea); MES (Latvia); LAS (Lithuania); MOE and UM (Malaysia); BUAP, CINVESTAV, CONACYT, LNS, SEP, and UASLP-FAI (Mexico); MOS (Montenegro); MBIE (New Zealand); PAEC (Pakistan); MES and NSC (Poland); FCT (Portugal); MESTD (Serbia); MCIN/AEI and PCTI (Spain); MOSTR (Sri Lanka); Swiss Funding Agencies (Switzerland); MST (Taipei); MHESI and NSTDA (Thailand); TUBITAK and TENMAK (Turkey); NASU (Ukraine); STFC (United Kingdom); DOE and NSF (USA).
  
  \hyphenation{Rachada-pisek} Individuals have received support from the Marie-Curie program and the European Research Council and Horizon 2020 Grant, contract Nos.\ 675440, 724704, 752730, 758316, 765710, 824093, 884104, and COST Action CA16108 (European Union); the Leventis Foundation; the Alfred P.\ Sloan Foundation; the Alexander von Humboldt Foundation; the Belgian Federal Science Policy Office; the Fonds pour la Formation \`a la Recherche dans l'Industrie et dans l'Agriculture (FRIA-Belgium); the Agentschap voor Innovatie door Wetenschap en Technologie (IWT-Belgium); the F.R.S.-FNRS and FWO (Belgium) under the ``Excellence of Science -- EOS" -- be.h project n.\ 30820817; the Beijing Municipal Science \& Technology Commission, No. Z191100007219010; the Ministry of Education, Youth and Sports (MEYS) of the Czech Republic; the Hellenic Foundation for Research and Innovation (HFRI), Project Number 2288 (Greece); the Deutsche Forschungsgemeinschaft (DFG), under Germany's Excellence Strategy -- EXC 2121 ``Quantum Universe" -- 390833306, and under project number 400140256 - GRK2497; the Hungarian Academy of Sciences, the New National Excellence Program - \'UNKP, the NKFIH research grants K 124845, K 124850, K 128713, K 128786, K 129058, K 131991, K 133046, K 138136, K 143460, K 143477, 2020-2.2.1-ED-2021-00181, and TKP2021-NKTA-64 (Hungary); the Council of Science and Industrial Research, India; the National Research Foundation of Korea (NRF/MSIT) grant No. 2020R1C1C1005916 (Korea); the Latvian Council of Science; the Ministry of Education and Science, project no. 2022/WK/14, and the National Science Center, contracts Opus 2021/41/B/ST2/01369 and 2021/43/B/ST2/01552 (Poland); the Funda\c{c}\~ao para a Ci\^encia e a Tecnologia, grant CEECIND/01334/2018 (Portugal); the National Priorities Research Program by Qatar National Research Fund; MCIN/AEI/10.13039/501100011033, ERDF ``a way of making Europe", and the Programa Estatal de Fomento de la Investigaci{\'o}n Cient{\'i}fica y T{\'e}cnica de Excelencia Mar\'{\i}a de Maeztu, grant MDM-2017-0765 and Programa Severo Ochoa del Principado de Asturias (Spain); the Chulalongkorn Academic into Its 2nd Century Project Advancement Project, and the National Science, Research and Innovation Fund via the Program Management Unit for Human Resources \& Institutional Development, Research and Innovation, grant B05F650021 (Thailand); the Kavli Foundation; the Nvidia Corporation; the SuperMicro Corporation; the Welch Foundation, contract C-1845; and the Weston Havens Foundation (USA).
\end{acknowledgments}

\bibliography{auto_generated} 

\cleardoublepage \appendix\section{The CMS Collaboration \label{app:collab}}\begin{sloppypar}\hyphenpenalty=5000\widowpenalty=500\clubpenalty=5000
\cmsinstitute{Yerevan Physics Institute, Yerevan, Armenia}
{\tolerance=6000
A.~Tumasyan\cmsAuthorMark{1}\cmsorcid{0009-0000-0684-6742}
\par}
\cmsinstitute{Institut f\"{u}r Hochenergiephysik, Vienna, Austria}
{\tolerance=6000
W.~Adam\cmsorcid{0000-0001-9099-4341}, J.W.~Andrejkovic, T.~Bergauer\cmsorcid{0000-0002-5786-0293}, S.~Chatterjee\cmsorcid{0000-0003-2660-0349}, K.~Damanakis\cmsorcid{0000-0001-5389-2872}, M.~Dragicevic\cmsorcid{0000-0003-1967-6783}, A.~Escalante~Del~Valle\cmsorcid{0000-0002-9702-6359}, P.S.~Hussain\cmsorcid{0000-0002-4825-5278}, M.~Jeitler\cmsAuthorMark{2}\cmsorcid{0000-0002-5141-9560}, N.~Krammer\cmsorcid{0000-0002-0548-0985}, L.~Lechner\cmsorcid{0000-0002-3065-1141}, D.~Liko\cmsorcid{0000-0002-3380-473X}, I.~Mikulec\cmsorcid{0000-0003-0385-2746}, P.~Paulitsch, J.~Schieck\cmsAuthorMark{2}\cmsorcid{0000-0002-1058-8093}, R.~Sch\"{o}fbeck\cmsorcid{0000-0002-2332-8784}, D.~Schwarz\cmsorcid{0000-0002-3821-7331}, M.~Sonawane\cmsorcid{0000-0003-0510-7010}, S.~Templ\cmsorcid{0000-0003-3137-5692}, W.~Waltenberger\cmsorcid{0000-0002-6215-7228}, C.-E.~Wulz\cmsAuthorMark{2}\cmsorcid{0000-0001-9226-5812}
% ...
\par}
\vskip\cmsinstskip
\dag:~Deceased\\
$^{1}$Also at Yerevan State University, Yerevan, Armenia\\
$^{2}$Also at TU Wien, Vienna, Austria\\
% ...
\end{sloppypar}
\end{document}